\documentclass[%
 reprint,
 amsmath,amssymb,
 aps,
]{revtex4-2}

\usepackage{graphicx}%
\usepackage{dcolumn}%
\usepackage{bm}%
\usepackage{hyperref}%
\hypersetup{colorlinks,bookmarksnumbered,
citecolor=cyan}

\usepackage{braket}
\usepackage{bm}
\usepackage{siunitx}
\usepackage{subcaption}
\usepackage{scalerel}
\usepackage{aas_macros}

\usepackage[usenames, dvipsnames]{xcolor}

\newcommand{\software}[1]{\texttt{#1}}

\renewcommand{\vec}[1]{\bm{#1}}

\newcommand{\msun}{\unit{M_\odot}}

\newcommand{\params}{\ensuremath{\theta}}

\newcommand{\hcomplex}{\ensuremath{\widetilde{h}(\vec{t}, \params{}_I)}}
\newcommand{\amp}{\ensuremath{A(\vec{t}, \freq{}, \chirpmass{})}}
\newcommand{\phase}{\ensuremath{\phi(\vec{t}, \freq{}, \chirpmass{})}}
\newcommand{\matern}{Matérn}
\newcommand*{\pdot}{\mathbin{\scalerel*{\boldsymbol\odot}{\circ}}}

\newcommand{\freq}{\ensuremath{f_I}}
\newcommand{\chirpmass}{\ensuremath{\mathcal{M}_c}}

\newsavebox{\largestimage}

\begin{document}

\preprint{APS/123-QED}

\title{A compact time-domain reduced order quadrature for gravitational wave inspirals with non-stationary noise}%

\author{Benjamin Zhang}
\email{zhangben@usc.edu}
\author{Kris Pardo}
\affiliation{University of Southern California, Los Angeles, CA 90089, USA}
\author{Bence Bécsy}
\affiliation{Institute for Gravitational Wave Astronomy and School of Physics and Astronomy, University of Birmingham, Edgbaston, Birmingham B15 2TT, UK}

\date{\today}%

\begin{abstract}
    We develop the first time-domain reduced order quadrature for gravitational waves in the early inspiral period, motivated by the need for dataset compression in the presence of gaps, uneven sampling, and dense covariance matrices.
    A single reduced basis with 1290 elements can represent inspirals from supermassive black hole binary sources with chirp masses between $10^7-10^{10}$ $M_\odot$, but in the presence of non-stationary noise $\sim 1290^2$ reduced order quadrature weights must be stored.
    We are able to reduce the stored weights by a factor of two by partitioning our parameter space by GW frequency, with relative waveform/likelihood accuracies remaining below $10^{-5}$/$10^{-3}$ after partition.
    Finally, we point out that heterodyning can be combined with reduced order methods with no loss in accuracy. 
    Our reduced order quadrature makes inspiral searches in the time domain feasible for both pulsar timing arrays and relative stellar astrometry.
\end{abstract}

\maketitle

\section{Introduction}

    When performing a search for a coherent gravitational wave (GW) signal, matched filtering against a bank of waveform templates provides the statistically optimal method.
    Early methods have therefore focused on constructing the optimal set of waveforms for a template bank using numerical relativity (NR) simulations \cite{Balasubramanian_1996_templatebank_infogeom, harry_2009_templatebank_stochastic, roulet_2019_templatebank}.
    However, the computational expense of NR waveform simulations, as well as the need to interpolate waveforms between templates in a bank for GW source parameter estimation, has led to various methods for producing fast and accurate waveform templates on demand.
    In particular, reduced order methods 
    aim to approximate a training set of waveforms by finding a subset that linearly spans the original set; in many cases, this subset is exponentially smaller
    \cite{tiglio_2022_ReducedOrderSurrogatea, field_2011_ReducedBasisCatalogs, field_2012_gw_rb}.
    This subset, or reduced basis (RB), can be combined with a nearly-optimal interpolation in physical space (either frequency or time) through the Empirical Interpolation Method (EIM), to create a waveform approximant which only needs to be exactly evaluated at $N$ physical points, where $N$ is the size of the reduced basis.
    These two techniques combined (``RB+EIM") allow the construction of surrogate models that are as accurate as NR simulations, at a fraction of the computational cost \cite{blackman_2015_surrogate, blackman_2017_surrogate, varma_2019_surrogate}.

    One important spinoff from reduced order modeling is the reduced order quadrature (ROQ) method.
    The ROQ method emerges from the RB+EIM approximant providing an affine parameterization of the waveform.
    This separates the waveform into a linear combination of $N$ basis vectors of length $P$ that do not depend on the GW parameters, and coefficients that do.
    Since the Gaussian log-likelihood typically used for Bayesian analyses is linear, noise-weighted inner products between the data and the basis vectors, and between the basis vectors with themselves, can be precomputed in an offline stage.
    This enables fast evaluation of the likelihood in an online stage, since the waveform has to be evaluated at only $N \ll P$ points \cite{antil_2013_TwoStepGreedyAlgorithm}.
    This has been used to speed up gravitational wave analyses for ground-based detectors by a factor of $\mathcal{O}(10-100)$ \cite{canizares_2015_AcceleratedGravitationalWave, smith_2016_FastAccurateInference}.

    In this work, we adapt reduced order quadrature methods for use in continuous GW Bayesian searches in relative stellar astrometry and pulsar timing arrays (PTAs).
    These methods are sensitive to GWs from supermassive black hole binaries in the ranges of $10^{-9}-10^{-4}\ \unit{\Hz}$ and $10^{-10}-10^{-6}\ \unit{\Hz}$ respectively \cite{wang_2021_GravitationalWaveDetection, zhang_2025_astrometricgw, nanograv_2026_15yr_cgw}.
    In astrometry, GWs at Earth manifest as apparent angular deflections of point sources; for PTAs, they manifest as pulsar timing delays that depend on the GW response at both the Earth and the pulsar.
    As reduced order methods were originally developed for waveforms detected in the LIGO/Virgo/KAGRA (LVK) frequency band, we deal with different considerations when adapting them to astrometry and PTAs.
    In both:
    \begin{enumerate}
        \item The waveform being modeled is generally much simpler than LVK binary merger waveforms. For astrometry, we expect sources up to $\sim 10^{-6}\ \unit{\Hz}$ to be in the early inspiral period. For PTAs, low-frequency sources below $\sim 10$ nHz can be well-approximated as a constant-frequency sinusoid, but at higher frequencies evolution between the Earth and pulsar terms will be measurable (see e.g.~Ref.~\cite{Petrov+2026}), and above $\sim 50$ nHz, the waveform can evolve appreciably even within the observing timespan \cite{sesana_2010_pta_freqevo}.
        \item Gaps in the data and uneven sampling mean continuous GW searches must be carried out in the time domain. %
        \item Residuals from complicated systematic effects make the noise non-stationary. For PTAs, these can come from modeling several non-GW processes that affect the timing of radio pulses, such as: pulsar spindown, pulse profile variability, changes in the interstellar medium resulting in dispersion and scattering, movement of the telescopes relative to the Solar System barycenter, etc. For astrometry, the primary concern is removing systematics associated with the telescope. As we show in Section \ref{sec:reduced-basis}, this complicates the construction of reduced order quadratures, as a separate ROQ cannot be constructed on the waveform magnitude $|h|^2$.
    \end{enumerate}

    For astrometry in particular, the size of our dataset, assuming a GW search using the \textit{Kepler Space Telescope}, is $\sim 200$ GB \cite{zhang_2025_astrometricgw}.
    This is due to the data having $\sim 10^5$ stars, with $\SI{6e4}{}$ observations per star (30 min cadence over 3.5 years, which are values we assume for the rest of this work). %
    This estimate also excludes the dense inverse covariance matrix per star expected from realistic treatments of systematics.
    More than speed of waveform evaluation, the overriding concern is reducing the size of this dataset before conducting our search.
    
    While the methods developed in this work are applicable to both PTAs and astrometry, we focus on the latter when detailing our formalism and numerical validation.
    
    Our paper proceeds as follows.
    We outline our GW parameter space, and summarize the principles behind reduced order quadratures.
    We discuss the complications caused by non-stationary noise, which leads to dense inverse covariance matrices, and describe a partitioning of the parameter space into overlapping regions that allows us to shrink the resulting reduced order quadrature.
    We numerically validate the accuracy of our reduced order quadratures; finally, we point out that reduced order quadratures can be exactly combined with heterodyning, another technique for computationally-efficient likelihood approximations.

\section{Reduced basis construction and partitioning}
\label{sec:reduced-basis}

    The log-likelihood for astrometry takes the following form:
    \begin{equation}
    \label{eq:astrometry-likelihood}
        \ln \mathcal{L} = -\frac12 (\vec{d} - \vec{s})^T C^{-1} (\vec{d} - \vec{s}) + \text{...}
    \end{equation}
    where the observed data $\vec{d}$ and candidate signal $\vec{s}$ depending on GW source parameters $\theta$ are concatenated across the observed x and y axes: $\vec{d} = [\vec{d}_x \ \vec{d}_y]^T, \vec{s} = [h_x(\vec{t}, \params{}) \ h_y(\vec{t}, \params{})]^T$.
    The omitted terms from Equation \ref{eq:astrometry-likelihood} are constant with respect to $\theta$, and can be discarded.
    Here, $C^{-1}$ is also a block inverse covariance matrix:
    \begin{equation}
        C^{-1} = 
        \begin{bmatrix}
            C_{xx} & C_{xy} \\
            C_{xy}^T & C_{yy}
        \end{bmatrix}^{-1}
        =
        \begin{bmatrix}
            A & B \\
            B^T & D
        \end{bmatrix}
    \end{equation}
    where $A, B, D$ are related to the original covariance blocks through the Schur complement.
    For convenience, from now on we abbreviate the inner product $\vec{a}^T M \vec{b}$ to $\braket{\vec{a}|M|\vec{b}}$.
    For a typical astrometric dataset, we observe each star for a few years with a cadence of $\sim \qty{10}{\min}$, so that even with diagonal covariances we must store $T \equiv \mathrm{len}(\vec{t}) \sim 10^{5}$ data points.
    
    In astrometry, both $h_x$ and $h_y$ are linear combinations of the $h_+$ and $h_\times$ polarizations that depend on the extrinsic parameters of inclination $i$, source sky position angle $(\phi_p, \theta_p)$, and polarization angle $\psi$ \cite{wang_2021_GravitationalWaveDetection}.
    We split the full parameter vector into intrinsic and extrinsic parameters, $\params{} = [\params{}_I \ \params{}_E]$.
    For a given fixed set of intrinsic parameters, if the inner products $\braket{\vec{d}_a | M | h_\alpha(\vec{t}, \params{}_I)}$ and $\braket{h_\alpha(\vec{t}, \params{}_I) | M | h_\beta (\vec{t}, \params{}_I)}$ are computed for combinations of $a \in [x, y], \alpha, \beta \in [+, \times], M \in [A, B, B^T, D]$, the dataset size for each star reduces from $10^5$ to $28$ points.
    When assuming a monochromatic waveform where the GW frequency $\freq{}$ is the only intrinsic parameter, the inner products can be gridded on $\freq{}$ and interpolated between grid points. This was developed for pulsar timing arrays in Ref. \cite{becsy_2024_quickcw2}, and adapted to astrometry in Ref. \cite{zhang_2025_astrometricgw}.

    We now want to extend this to early inspiral waveforms with intrinsic parameters $\params{}_I = [\freq{}, \chirpmass{}, \delta]$.
    Here, $\freq{}$ and $\delta$ are defined as the instantaneous frequency and global phase respectively of the GW when it enters our observation period.
    The chirp mass $\chirpmass{}$ determines the frequency evolution of an inspiralling waveform.
    If we construct a RB and accompanying EIM points, we obtain an affine parameterization on the waveform:
    \begin{equation}
    \label{eq:affine-parameterization-real}
        h_{\alpha} (\vec{t}, \params{}_I) \simeq \sum_{a=1}^N B_a(\vec{t}) h_\alpha (T_a, \params{}_I) 
    \end{equation}
    where $T_a$ is the fixed EIM time associated with EIM basis vector $B_a(\vec{t})$.
    The previously-mentioned inner products can then be further broken up into linear combinations of ROQ weights and waveform evaluations at $T_a$.
    The ROQ weights are inner products between combinations of the data and EIM basis vectors.
    
    To do this, we have a choice: we can build a ROQ directly on the real waveform $h_+(\vec{t}, \params{}_I) = \amp{} \cos{(\phase{} + \delta)}$, gridding across $\delta$ from 0 to $2\pi$.
    This covers $h_\times$ as well, because it follows the same waveform as $h_+$, only shifted by $\pi/2$: $h_\times \propto \sin \phi = \cos(\phi - \frac\pi2)$.
    Alternatively, we can build a ROQ on the complex waveform $\hcomplex{} = h_+ + ih_\times = \amp{} \exp{(i \phase{})} \exp{(i \delta})$, and use the formulas $h_+ = \Re{[\widetilde{h}]} = \frac12 (\widetilde{h} + \widetilde{h}^*)$ and $h_\times = \Im{[\widetilde{h}]} = \frac{1}{2i} (\widetilde{h} -  \widetilde{h}^*)$. 
    Numerically, we find that when taking into account the doubled memory required to store a complex ROQ weight, the sizes of ROQs built using either choice are equal.
    However, we opt for the latter choice,
    because $\delta$ then factors out of the EIM basis vectors $B_a(\vec{t})$.
    We can also omit gridding the reduced basis training set over different values of $\delta$, which saves a significant amount of memory during ROQ construction.

\begin{figure*}
        \centering
        \includegraphics[width=0.6\linewidth]{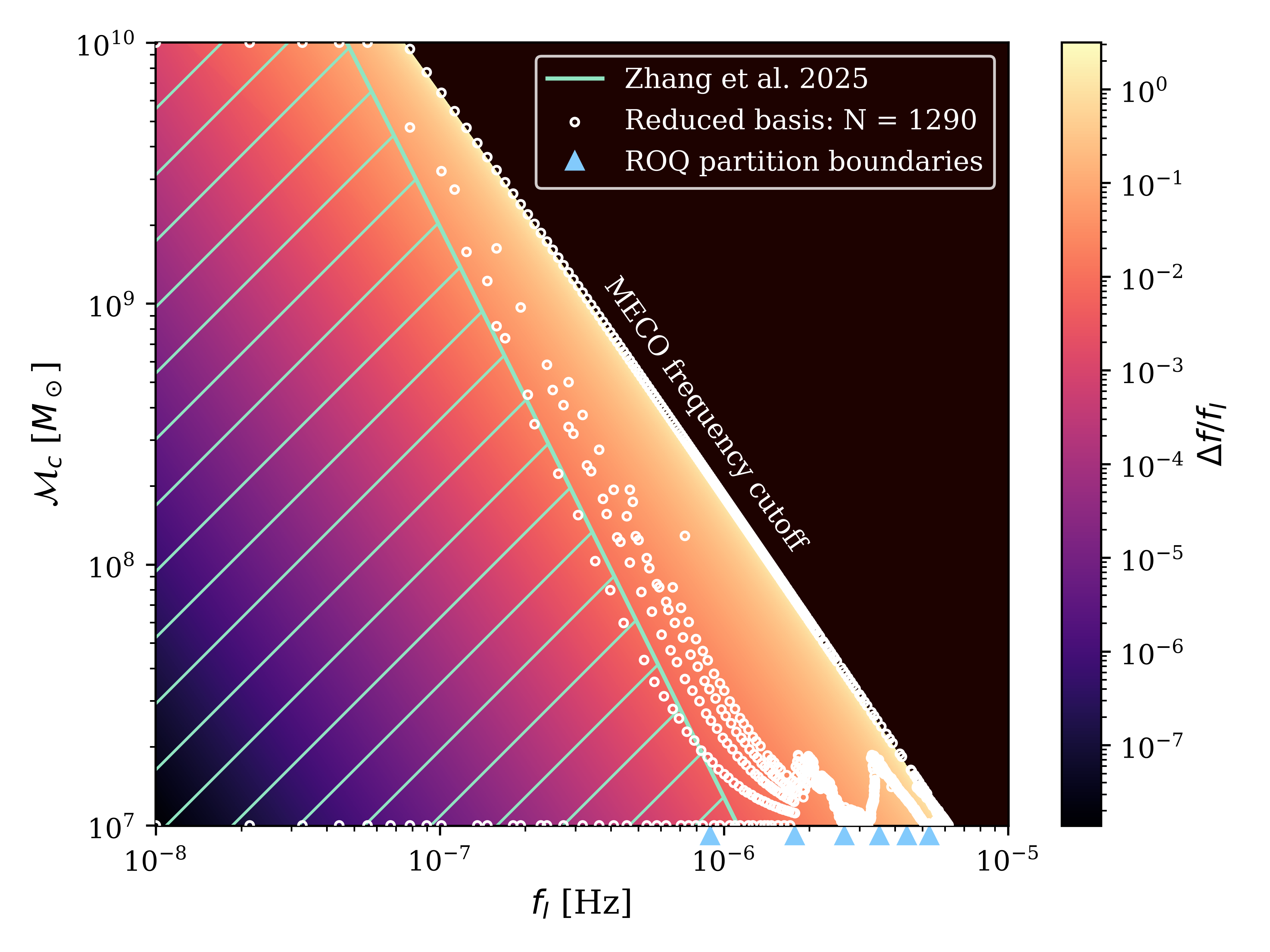}
        \caption{The inspiral parameter space used to build our time-domain reduced basis, and the elements chosen for the basis. The parameter space is colored by the relative change in frequency over a 3.5 year observation period, and is cut off when the maximum frequency during observation reaches the minimum energy circular orbit (MECO) frequency. This limits the reduced basis to early inspirals. The method presented in this paper expands upon the previous region of validity used for Ref. \cite{zhang_2025_astrometricgw}, shown in light teal, which assumes approximately constant GW frequency. The training set waveforms chosen for the full reduced basis are the white hollow circles. The apparent structure in the chosen waveforms at $\freq{} < 10^{-6} \ \unit{\Hz}$ is an artifact of the training set gridding. The frequency boundaries for our partitioned reduced order quadrature are marked with light blue arrows.}
        \label{fig:training-param-space}
    \end{figure*}

    To generate our time-domain waveform amplitude, $\amp{}$, and phase, $\phase{}$, we use the \software{IMRPhenomT} phenomenological model, which accurately represents the dominant $(2, 2)$ mode for quasicircular, non-precessing binary black holes \cite{estelles_2021_imrphenomt}. We use the implementation in the \software{phenomxpy} library \cite{garciaquiros_2025_phenomxpy}.
    We restrict ourselves to an equal mass ratio $q = 1$, and no spin. 
    In addition, to restrict ourselves to the early inspiral period, we only consider waveforms that enter the observation period at a frequency $\freq{}$ and leave the observation period at a final frequency less than the minimum-energy circular orbit (MECO) frequency $f_\text{MECO}$.
    $f_\text{MECO}$ is often taken to be the transition between the early inspiral and late inspiral/intermediate regime for phenomenological models.
    Instead of defining the GW phase with the boundary condition $\phi(t_\text{coalescence}) = 0$, we mandate that $\phi(t_0) = \delta$, where $t_0$ is the first observed time. As previously mentioned, for the reduced basis training set we can set $\delta = 0$ with no accuracy loss.
    We additionally remove waveforms that reach a maximum frequency greater than our Nyquist frequency of $({2 \times 30 \ \unit{min}})^{-1}$.
    Finally, we only consider chirp masses between $10^7 \ \msun{}$ and $10^{10} \ \msun{}$.
    Our allowed parameter space is shown in Figure \ref{fig:training-param-space}.

    We generate our training set waveforms by gridding linearly on initial frequency assuming the minimum chirp mass, and for each selected initial frequency gridding linearly on chirp mass, up to the maximum chirp mass allowed by the MECO frequency condition.
    We set the grid spacing empirically, based on numerical convergence of the number of reduced basis elements.

    We use the \software{arby} code to generate a greedy reduced basis + EIM (RB+EIM) approximant on our training set with a greedy error threshold of $10^{-12}$ \cite{villanueva_2021_Arby}.
    The full reduced basis has 1290 elements, which are plotted in Figure \ref{fig:training-param-space}.
    The picked elements are largely clustered on the high-frequency end and along the edges of the allowed parameter space, a phenomenon also seen in other GW reduced bases \cite{field_2012_gw_rb}.
    Although the picked elements are sparse on the low-frequency end, we find from numerical validation (Section \ref{sec:rb-validation}) that the RB+EIM waveform remains accurate there.
    Since our training set waveforms are complex, the EIM basis vectors in our affine parameterization are complex as well:
    \begin{equation}
        \widetilde{h}_{\alpha}(\vec{t}, \params{}_I) \simeq \sum_{a=1}^N \widetilde{B}_a(\vec{t}) \widetilde{h}_\alpha (T_a, \params{}_I) 
    \end{equation}

    Unlike previous studies of GW reduced bases, we are unable to assume stationary noise in the time-domain, which translates to a diagonal covariance matrix in the frequency domain.
    Having a diagonal covariance matrix allows a separate reduced basis to be built on the waveform magnitude $|h(\vec{f}, \params{}_I)|^2$, but we cannot do this for arbitrary dense inverse covariance matrices $M^{(i)} \in [A, B, B^T, D]$.
    As such, when constructing a set of ROQ weights from our RB+EIM affine parameterization, we must store both the linear weights $\rho_{a}^{(i)} \equiv \braket{\vec{d}|M^{(i)}|\widetilde{B}_a(\vec{t})}$ and the `quadratic weight matrices' $\xi^{(i)}_{ab} \equiv \braket{\widetilde{B}_a(\vec{t})|M^{(i)}|\widetilde{B}_b(\vec{t})}$, $\omega^{(i)}_{ab} \equiv \braket{\widetilde{B}_a(\vec{t})^*|M^{(i)}|\widetilde{B}_b(\vec{t})}$. 
    The total size of our ROQ weights is dominated by the quadratic weight matrices:
    \begin{equation}
        N_\text{ROQ} \propto 2N^2 + N = N(2N + 1)
    \end{equation}
    where $N$ is the number of elements in our reduced basis.
    Depending on how dense our $T \times T$ inverse covariance matrices $M^{(i)}$ are, constructing an ROQ directly on our full reduced basis still results in a dataset compression factor of $\mathcal{O}(1000)$ in the fully-dense case, with $T \sim 60000$.
    We can improve on this by dividing our parameter space into partitions, and constructing a separate reduced basis for each partition.
    This has been done before for gravitational waves in the LIGO-Virgo-KAGRA (LVK) frequency range to shrink the total size of the reduced bases \cite{smith_2016_FastAccurateInference,  morisaki_2020_RapidParameterEstimation, cerino_2023_AutomatedParameterDomain}.
    In our case, we find the total number of reduced basis elements summed across all partitions is greater than the size of the original reduced basis.
    However, since our ROQ size is dominated by the quadratic weight matrices, this can still lead to a smaller total size:
    \begin{equation}
        N_\text{ROQ,partitioned} \propto \sum_{i=1}^{N_\text{part}} N_i (2N_i + 1)
    \end{equation}

    We partition on just the initial GW frequency $\freq{}$ for simplicity, and to aid in stitching the partitions back together in the full likelihood, as detailed later.
    We minimize the ratio between $N_\text{ROQ,partitioned}$ and $N_\text{ROQ}$, the `reduction factor', exploring two strategies.
    We either split the parameter space into partitions that span an equal width in $\freq{}$ at minimum allowed $\chirpmass{}$, or split so that the number of training set waveforms in each partition is approximately equal.
    We evaluate these strategies for $N_\text{part} = [2, 10]$, and find that the best-performing strategy is to split into 7 equal-$\freq{}$-width segments (whose boundaries are marked in Figure \ref{fig:training-param-space}).
    Under this partitioning scheme, our reduction factor is 0.53.
    Interestingly, the reduction factor stays almost constant for a wide range of partition numbers; for this strategy, splitting into $4-10$ partitions yields a reduction factor between $0.53$ and $0.56$, which only increases to $0.59$ and $0.69$ for 3 and 2 partitions respectively.
    This may indicate that our partition scheme is not optimal, which we aim to follow up on in future work.

    If we choose which set of ROQ weights to use based on the candidate waveform's $\freq{}$, the waveform may have sudden discontinuities when crossing from one partition to another.
    To mitigate this, when we partition our parameter space, we have the partitions overlap in $\freq{}$.
    Since the reduced basis construction process terminates when the maximum projection error across all training set waveforms is below a threshold, this guarantees that any discontinuities when moving between partitions are comparable to those within a single partition.
    We choose to overlap by 2 frequency grid points on each side of the boundary frequency between a pair of partitions.
    In Section \ref{sec:rb-validation}, we numerically test whether this overlap is sufficient.

\section{Numerical validation of likelihood accuracy}
\label{sec:rb-validation}

    To validate the accuracy of our partitioned ROQs, we first calculate the $L_\infty$ error between the RB+EIM approximated waveform and the exact waveform.
    This is the maximum absolute difference across all time points between the two waveforms; since both waveforms are normalized, this can also be approximately interpreted as a maximum relative error.
    We calculate the error for a set of validation waveforms generated using the same gridding procedure described in Section \ref{sec:reduced-basis}, but with twice the number of grid points in both $\freq{}$ and $\chirpmass{}$.
    Due to our training set construction process, no waveforms in the validation set are in the training set.
    For each partition, we validate on all waveforms that lie within the partition's nominal frequency bounds (i.e. not counting the overlaps between partitions).
    In our chosen partitioning scheme, the maximum $L_\infty$ validation error across all partitions is $\qty{4.81e-4}{}$, with a mean $L_\infty$ validation error (combined across partitions) of $\qty{1.66e-5}{}$.

    We also test the accuracy of our partitioned ROQ likelihood on a toy time-domain problem meant to replicate aspects of an astrometric GW search.
    For this problem, we assume all extrinsic parameters are fixed so that the data contains only the $h_+$ polarization, on one astrometric spatial axis.
    We use the same evenly-spaced time grid used for construction of our reduced basis, but due to computational limits, we downscale the time grid by $6 \times$ by taking evenly-spaced subsamples.
    The RB+EIM approximants that we constructed in the previous section for each partition can still be used by subsampling the EIM basis vectors in the same way.
    This downsampling may lead to us underestimating the ROQ error, but we do not expect this to significantly change our results. 
    
    We construct a toy covariance consisting of a white noise diagonal of ones, added to a \matern{} component with a length scale of 90 days and smoothness parameter $\nu = 3/2$.
    Before adding the two components, we scale down the \matern{} component by $100 \times$ (so that the final diagonal of the covariance is $1.01$). 
    While the \matern{} covariance is stationary, we choose this length scale to emulate the case in astrometry where we have residuals from uncleaned systematics that produce low-frequency correlations in the data, and therefore a dense inverse covariance.
    In addition, in real data, gaps and uneven sampling prevent us from transforming into the frequency domain to diagonalize a stationary covariance.

    We inject a mock inspiral $h_+$ signal into 100 noise realizations of the covariance, at the parameters $(\freq{} = \SI{1.04e-6}{\Hz}, \chirpmass{} = \SI{3e7}{\msun{}}, \delta = \frac\pi2)$.
    For astrometry, this is akin to having 100 stars at the same on-sky position.
    We choose the injection frequency to coincide with the border between the first two ROQ partitions, to validate that the likelihood does not significantly change between partitions.
    We scale the amplitude of the waveform such that the signal-to-noise ratio, assuming white noise, is 3.5.
    Precisely, we normalize the waveform by dividing it by its average magnitude, then scaling by another factor of $(2 \sigma_w T N_s)^{-1/2}$, where $\sigma_w$ is the diagonal of the total covariance matrix, $T$ is the number of observations, and $N_s$ the number of stars.
    We choose to inject the same signal into multiple noise realizations since our covariance means low-frequency noise dominates for any given realization, leading to spurious likelihood peaks.
    In reality, truly independent red noise for each star will mitigate these peaks (although shared red noise from telescope systematics can lead to the opposite effect).
    Since our SNR scaling factor only assumes white noise, going from one noise realization to many also degrades our overall signal recoverability, as each new realization adds additional non-white noise.

    We only sample over the frequency $\freq{}$ and chirp mass $\chirpmass{}$, fixing all other parameters (waveform amplitude and initial phase) to their true values. 
    We impose a uniform prior on the joint density of $\log_{10} \freq{}$ and $\log_{10} \chirpmass{}$ within the region where the reduced basis is valid. 
    This prior almost certainly overweights the low frequency end, as more of the $\chirpmass{}$ space is available there, but we believe this is acceptable for our purposes.

\begin{figure*}
        \centering
        \savebox{\largestimage}{\includegraphics[width=0.45\linewidth]{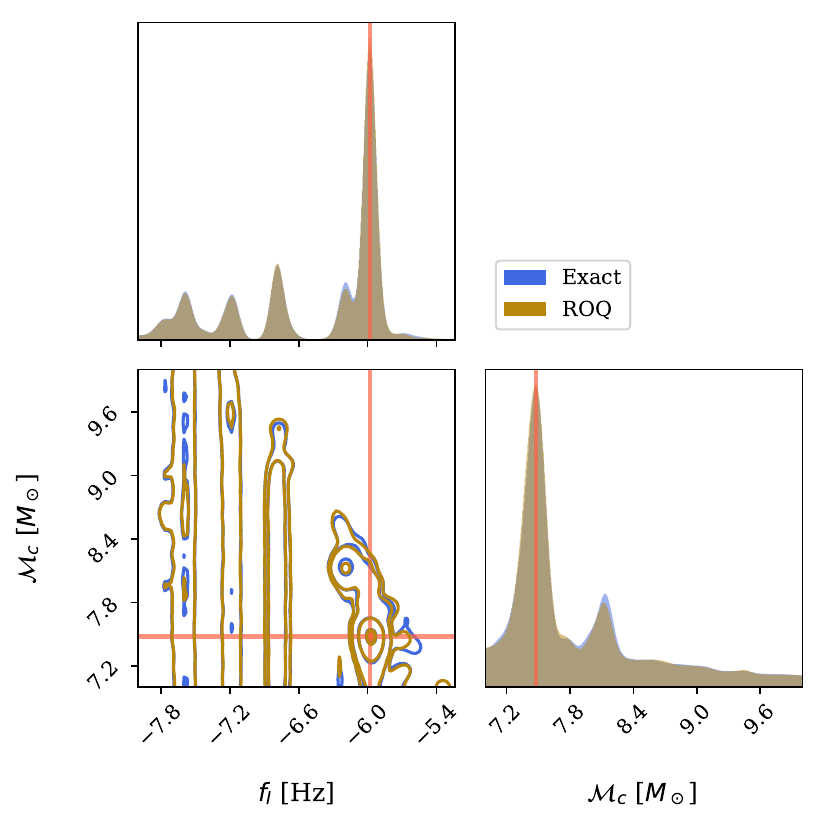}}%
        \begin{subfigure}{.5\textwidth}
            \centering
            \usebox{\largestimage}
        \end{subfigure}%
        \begin{subfigure}{.5\textwidth}
            \centering
            \raisebox{\dimexpr.5\ht\largestimage-.5\height}{%
                  \includegraphics[width=\linewidth]{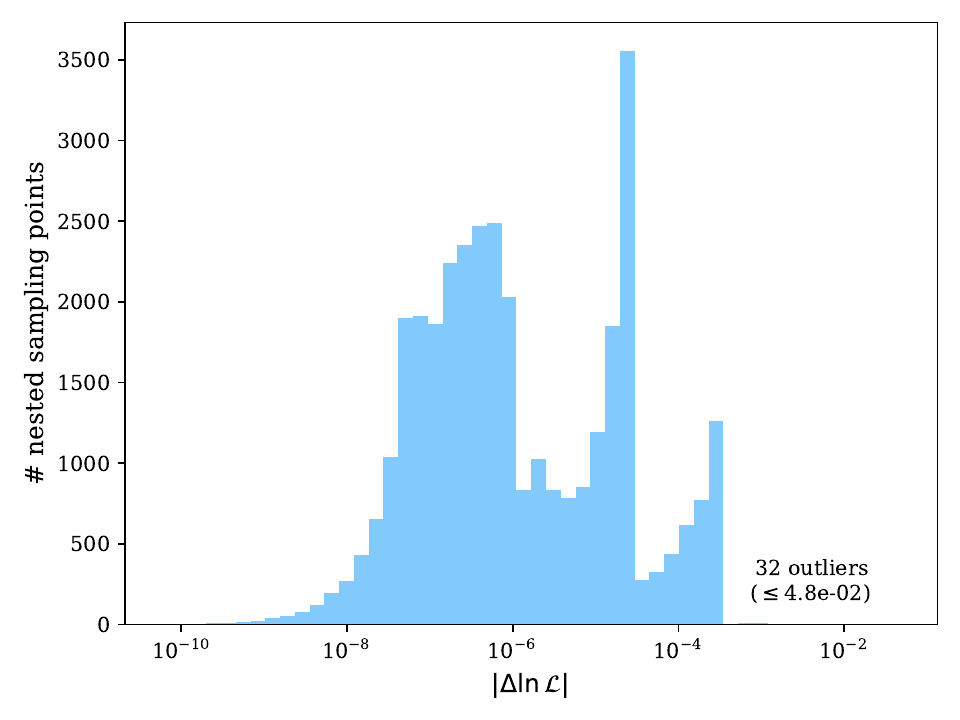}}
        \end{subfigure}
        \caption{\textit{Left}: nested sampling posterior with an exact likelihood (blue) vs. a likelihood using reduced order quadratures (gold). The true parameters of the injected signal are shown in red. Despite both nested sampling runs being started with the same random state and sampling parameters, small differences in the two likelihoods lead to slightly different posteriors. The low-frequency spikes are caused by the off-diagonal \matern{} covariance, which adds red noise to the data. \textit{Right}: The absolute log-likelihood error, calculated for each sampled point in the exact nested sampling run. The close agreement between the two indicates that posterior differences for the nested sampling run likely stem from cascading divergences. Note that while the posterior on left is calculated from the samples weighted by their evidence contribution, the points shown in this histogram have no additional weighting.}
        \label{fig:rb-accuracy-validation}
    \end{figure*}

    Using both the exact and ROQ-based likelihoods, we obtain posteriors on $\freq{}$ and $\chirpmass{}$ using a static nested sampler from \software{dynesty} \cite{speagle_2020_dynesty} with the same settings and random state.
    Our posteriors are shown in the left panel of Figure \ref{fig:rb-accuracy-validation}.
    The dependence of nested sampling-derived posteriors on relatively few highly-weighted samples may lead to the small differences between the exact and ROQ likelihood shown here, but overall the posteriors are similar.

    For all samples drawn during the exact-likelihood nested sampling run, we calculate the ROQ likelihood at that sample's parameters and show the absolute difference in log-likelihood between the exact likelihood in the right panel of Figure \ref{fig:rb-accuracy-validation}.
    This difference is below $\SI{3.5e-4}{}$ for almost all samples (with 32 outliers below $\SI{4.8e-2}{}$), indicating extremely close agreement in both the low and high-likelihood regimes explored by the sampler.

\section{Heterodyning reduced bases}
\label{sec:local-heterodyning}

    Finally, we make the following observation about reduced order methods and heterodyning.
    
    First, we briefly review the usage of heterodyning (also referred to as `relative binning') for gravitational wave likelihoods.
    First proposed in Ref.~\cite{cornish_2010_FastFisherMatrices}, heterodyning uses the fact that the ratio between a candidate waveform and a reference waveform $H^\text{(r)}(t) = A^\text{(r)}(t) \exp{(i \phi^\text{(r)}(t))}$ in the parameter neighborhood of the reference waveform is slowly changing with respect to time (or frequency), and so can be approximated by a linear interpolation of the exact waveform evaluated at a sparse set of time bins multiplied by precomputed ``summary data".
    While originally used for frequency-domain analyses, \cite{cornish_2021_HeterodynedLikelihoodRapid, zackay_2018_relbinning} heterodyning has been adapted for the time domain as well \cite{sharma_2026_RapidInferenceGravitationalwave}.

    The reference waveform divided by itself can be inserted into the likelihood, so that the data-waveform inner product term of the likelihood becomes
    \begin{equation}
        \braket{\vec{d}|h(\vec{t}, \params{})} = \braket{\vec{d} \pdot{} H^\text{(r)}(\vec{t}) | \frac1{H^\text{(r)}(\vec{t})} \pdot{} h(\vec{t}, \params{})}
    \end{equation}
    where here we have elided the inverse covariance matrix, and $\pdot{}$ is the elementwise product.
    If we construct an affine parameterization for the ratio waveform $r(\vec{t}, \params{}) \equiv (1/{H^\text{(r)}(\vec{t})}) \pdot{} h(\vec{t}, \params{})$, the now-quickly-varying data term $\vec{d} \pdot{} H^\text{(r)}(\vec{t})$ is absorbed into the ROQ weights during ROQ construction.
    Ignoring numerical precision, this is \textit{exact}, and under a similar line of reasoning the quickly varying term for the waveform-waveform inner product can also be absorbed into a set of ROQ quadratic weight matrices.
    Compared to the cost of constructing a reduced basis, the computational cost of this additional elementwise product is trivial.

    Why would we want to do this?
    As in the heterodyned likelihood approximation, if the ratio waveform has fewer cycles than the original waveform, it may yield a smaller reduced basis and thus fewer ROQ elements.
    Unfortunately for the inspirals we consider, heterodyning all training set waveforms in an initial frequency partition by a reference waveform in the partition (for example, the smallest $\freq{}$ \& $\chirpmass{}$ waveform) does not change the number of reduced basis elements picked.
    This may be due to waveforms being extremely well-sampled even at the highest frequencies we consider, but finding the precise reason would be a good avenue for future work.
    Combining heterodyning with reduced bases may be more useful for LVK-band frequency-domain waveforms, as studies suggest that the size of reduced bases for these waveforms go up as the frequency spacing decreases \cite{morras_2023_eigenroq}.
    This may also allow for the further shrinkage of reduced bases by expanding the validity range of the multibanding method, where waveforms can be downsampled in the frequency ranges where they oscillate more slowly. 
    In Ref.~\cite{newell_2026_MultibandedReducedOrder}, the reduced basis size goes down after multibanding is applied, which may indicate support for a link between basis size and sampling rate.

\section{Conclusion}

    Motivated by the large dataset sizes expected from searches for continuous gravitational waves with relative astrometry and future PTA surveys like DSA-2000 and SKA \cite{hallinan_2019_dsa2000, wang_2021_skapta}, we develop the first reduced order quadrature (ROQ) for inspiralling gravitational waves in the time domain.
    In the presence of non-stationary Gaussian noise, the size of our reduced order quadrature is dominated by the waveform-waveform component of the likelihood, so that a reduced basis with $N$ elements produces an ROQ $\propto N^2$.
    We are able to shrink the quadrature size by a factor of two by partitioning the reduced basis by GW frequency, and validate that this partitioned ROQ remains accurate both in the waveform, and the associated likelihood.
    Finally, we make the to our knowledge novel observation that heterodyning can be exactly combined with reduced order quadratures, allowing a reduced basis to be built on the slowly-changing ratio waveform.
    This does not change the reduced basis size for the inspirals we consider, but may for waveforms in the LIGO-Virgo-KAGRA (LVK) frequency band.

    The toy Bayesian search we use to validate the accuracy of our ROQ likelihood is sufficient for this purpose, but is admittedly unrealistic; choices of covariance and prior aside, we assume the extrinsic parameters of the GW (such as sky position and polarization) are fixed so that the signal only appears in the $h_+$ polarization on one spatial axis.
    In follow-up work, we plan to fully implement the ROQ likelihood for intrinsic and extrinsic GW parameters in a realistic astrometric search.
    In addition, we plan to apply reduced order methods to the pulsar timing array band.
    While waveforms for pulsar timing arrays on the low-frequency end justifiably assume no frequency evolution within the observing timespan, the large non-stationarity of noise sources means that template-based searches may miss possible extremely-high-mass binaries and sources on the high-frequency end that exhibit significant increases in frequency. In addition, eccentric binaries can have faster frequency evolution and more complex waveforms (see e.g.~Ref.~\cite{Taylor+2016}), which could benefit from the treatment presented here, even at the lower frequencies probed by PTAs.

    With recent works exploring non-stationary noise in the LVK band \cite{cornish_2026_NonstationaryNoiseGravitational, cornish_2026_ModelingNonstationaryNoisea}, the extension of reduced order quadratures to non-stationary noise presented here could be applicable in either the time or frequency domain to LVK waveforms.
    The quadratic ROQ weight matrices needed for a non-diagonal covariance reuse the same reduced basis that is constructed for the waveform.
    Therefore, if the speed of online likelihood evaluations is dominated by evaluating the exact waveform coefficients $h(T_a, \theta_I)$ in Eq. \ref{eq:affine-parameterization-real}, the computational cost of evaluating the likelihood for non-stationary noise is nearly the same as for stationary noise.

\begin{acknowledgments}

    We are grateful for valuable discussions with Aaron Johnson, Katerina Chatziioannou, and Patrick Meyers during the course of this work. BZ and KP gratefully acknowledge support from the W.M. Keck Foundation Bridge Funding program.

\end{acknowledgments}

\bibliography{main}%

\end{document}